\documentclass[11pt,english,fleqn]{extarticle}
\usepackage{mathpazo}
\usepackage[scaled=0.92]{helvet}
\usepackage{courier}
\usepackage[T1]{fontenc}
\usepackage[latin9]{inputenc}
\usepackage{geometry}
\usepackage{babel}
\usepackage{array}
\usepackage{rotating}
\usepackage{float}
\usepackage{units}
\usepackage{multirow}
\usepackage{amsmath}
\usepackage{amsthm}
\usepackage{graphicx}
\usepackage{setspace}
\usepackage[authoryear]{natbib}
\usepackage[unicode=true,pdfusetitle,
 bookmarks=true,bookmarksnumbered=false,bookmarksopen=false,
 breaklinks=false,pdfborder={0 0 1},backref=false,colorlinks=false]
 {hyperref}
\makeatletter
\numberwithin{equation}{section}
\numberwithin{figure}{section}
\date{}
\usepackage[mathlines,displaymath]{lineno}
\usepackage[ruled]{algorithm2e}
\usepackage{authblk}

\@ifundefined{showcaptionsetup}{}{%
 \PassOptionsToPackage{caption=false}{subfig}}
\usepackage{subfig}
\makeatother

\begin{document}
\title{Dynamic Mixture of Experts Models for Online Prediction}
\author{Parfait Munezero\thanks{Department of Statistics, Stockholm University, Stockholm, Sweden
and Data Insights Support Team, Ericsson, Stockholm, Sweden (Corresponding
author: parfait.munezero@ericsson.com).}\,, \,Mattias Villani\thanks{Department of Statistics, Stockholm University, Stockholm, Sweden
and Department of Computer and Information Science, Link\"{o}ping University,
Link\"{o}ping, Sweden.} \,\,and Robert Kohn\thanks{UNSW Business School, University of New South Wales, Sydney, Australia.}}
\maketitle
\begin{abstract}
A mixture of experts models the conditional density of a response
variable using a mixture of regression models with covariate-dependent
mixture weights. We extend the finite mixture of experts model by allowing the parameters in
both the mixture components and the weights to evolve in time by following
random walk processes. Inference for time-varying parameters in richly
parameterized mixture of experts models is challenging. We propose
a sequential Monte Carlo algorithm for online inference and based
on a tailored proposal distribution built on ideas from linear Bayes
methods and the EM algorithm. The method gives a unified treatment
for mixtures with time-varying parameters, including the
special case of static parameters. We assess the properties of the
method on simulated data and on industrial data where the aim is to
predict software faults in a continuously upgraded large-scale software
project.
\end{abstract}

{\it Keywords}: Bayesian sequential inference, Linear Bayes, Particle filtering, Mixture models, Sequential Monte Carlo.

\section{Introduction}

A mixture of experts (ME) model \citep{jordan1994hierarchical} provides a flexible framework for expressing the distribution of a response
variable conditional on a set of covariates. It models continuous or discrete response variables using a finite mixture model with covariate-dependent component models and mixture weights. The component models are commonly referred to as the experts and the mixture weights models as gates. Generally, the experts are regression models with the response's conditional density from the exponential family; see \citet{gormley2018mixtures} for a concise introduction to ME models and \citet[Section IX]{yuksel2012twenty}  for an extensive list of application areas.

ME models have been extended in many ways. \citet{hunter2012semiparametric} relax the parametric assumption of the component models and propose a semi-parametric inference methodology for mixtures of linear regression models.  \citet{villani2012generalized} extend the component models to density functions outside the exponential family and use Bayesian variable selection in all parts of the model. \citet{jacobs1997bayesian} propose a Bayesian hierarchical model in which the conditional density of the response is expressed as a mixture of ME components. \citet{rasmussen2002infinite} propose an infinite mixture of Gaussian process experts. \citet{zeevi1996time} and \citet{carvalho2005mixtures, carvalho2005modeling} extend the ME model framework to autoregressive time series data where the covariates may include lagged values of the response and they study the properties of the maximum likelihood estimator. \citet{wood2002bayesian} propose a Bayesian inference methodology for ME models and allow smoothing spline components. In longitudinal data applications, \citet{muthen1999finite} consider fixed and random effect covariates in both components and mixture weights. \citet{quiroz2013dynamic} propose a ME model with static parameters for longitudinal data but where the subjects are allowed to dynamically change mixture components over time. 

While many existing ME models discussed in the previous paragraph are very flexible, they are still too restrictive for many applications, in particular when: i) the data is an irregularly spaced time series or arrives in batches; ii) the conditional density of the response tends to change over time. Our application to predicting software faults, in Section \ref{subsec:Software-trouble-reports}, is one such typical example taken from industry. Software is generally upgraded at irregular times - a week may pass without any release followed by two or three releases in the next week - depending on the bugs reported in previous versions, the complexity of the added features in the release, or business related factors. Also, the distribution of the response variable naturally changes as the software matures: the developers in the project change over time, user behavior changes, and new technologies emerge which demand adaptation.  All these issues make standard ME models for time series data impractical.

We extend the class of ME models to  dynamic mixture of experts (DME) models to propose an online (real-time) predictive model with time-varying parameters. The new class is particularly suited for unstructured streaming data, but may also be used for equally spaced time series with time-varying distributions. The proposed DME models have the general form of varying-regression coefficient models \citep{hastie1993varying}; the regression coefficients in both the mixture components and the mixture weights are allowed to vary over time through a latent process. The latent process can be modelled  as a discrete first order Markov process with static \citep{liehr1999hidden, kohlmorgen2000identification} or time-varying \citep{wang2003time} transition probabilities. Alternatively, it can be modelled through random walk processes \citep{west1985dynamic,fahrmeir2004penalized,fahrmeir2011bayesian} as in Section \ref{subsec:Prior-process}. The  density for the response variable is modelled conditional on the value of the latent state. This allows the model to adapt locally to any abrupt changes through time \citep{fahrmeir2004penalized}. Also, it facilitates designing  an adaptive online (real-time) Bayesian  predictive inference in which the uncertainty is updated sequentially as new data arrive and  predictions are computed based on the recently updated posterior density - the $online$ posterior; see Section \ref{sec:Inference}.

 Inference in mixture models is challenging. One challenge is  the identifiability issue caused by the invariance of the likelihood under permutation of component labels and parameters.  Constraints on the regression coefficients have been imposed to identify ME models, see \citep{jiang1999identifiability} and Appendix \ref{sec:identifiability}. However, identifiability is of less importance in applications where prediction is the main objective as the predictive distribution is unaffected by label switching \citep{geweke2007interpretation}. Another important issue in mixture models is the number of components. One strand of the literature use nonparametric Bayes approaches with infinite mixtures typically modeled by stick-breaking processes \citep{hjort2010bayesian}, or reversible jump MCMC \citep{richardson1997bayesian} to obtain the posterior over the number of mixture components. Another line of research uses model comparison methods to select the optimal number of components \citep{celeux2019model, geweke2007smoothly}. In addition to these issues, the structure of DME models brings additional challenges. DME are often richly parameterized with time-varying regression coefficients in both the components and the mixture weights. This leads to a high dimensional and complex target posterior that can only be properly explored by carefully designed numerical methods.
 
 We propose online inference based on Sequential Monte Carlo (SMC) methods \citep{del2006sequential,doucet2000sequential} to address these issues. The performance of SMC methods depends on the proposal/importance distribution from which parameters are sampled. Given the complexity of DME models, off-the-shelf SMC methods based on commonly used proposal distributions such as the bootstrap filter will perform poorly.  Our main contribution is a proposal distribution tailored to the class of DME models. The potentially high-dimensional regression coefficients influence the conditional density only through the low-dimensional linear predictors transformed by the link functions. This makes it possible to combine the linear Bayes method \citep{west1985dynamic} and ideas from the expectation and maximization (EM) algorithm principle \citep{dempster1977maximum}  to build  the proposal distribution efficiently.  Our algorithm builds on the marginal particle filter algorithm \citep{klaas2012toward} for online (real-time) prediction; offline inference using SMC algorithms can be done following ideas in \citet{munezero2021efficient}, where the proposed methodology can be considered as an integrated part of the  particle smoother.

Standard inference for mixture models requires posterior draws of component allocation indicators or, in the approach of \citet{carvalho2010particle}, to keep track of the number of allocations for each component and conditional sufficient statistics for the mixture component parameters. Our methodology expresses the likelihood in its marginal form, which does not require sampling component indicators, hence reducing  considerably the dimension of the target posterior.  The posterior's dimension is reduced even further by using the discount factor approach \citep{west1985dynamic} that recursively  estimates the innovation variance of the states at a particular time point as a function of a discount factor $0<\alpha<1$ and  the filtered information from the most recent posterior. This means that the inference  requires  keeping track of only the regression coefficients through time. 

The smoothness of the regression coefficients evolution is controlled via $\alpha$  which allows building static and dynamic models in a unified way just by changing the value of $\alpha$. \citet{liu2001combined} suggest that models with $0.95\leq \alpha < 1$  are essentially static, and those with $\alpha < 0.95$ are dynamic, although this rule of thumb may vary depending on applications. Inference on the discount factor and the number of components is performed using the log predictive score, a marginal likelihood-based model comparison criteria \citep{geweke2007smoothly,villani2009regression}.

The proposed methodology allows writing general computer code where a user can easily add a new model by supplying the first and second derivative
of the component densities with respect to the linear predictors, which are the only key arguments of our procedure. Recent advances in automatic differentiation \citep{baydin2018automatic} even removes the requirement of computing derivatives analytically.

The rest of the paper is organized as follows. Section \ref{sec:The model} introduces the dynamic mixture of experts model and the prior process. Section \ref{sec:Inference} presents the SMC algorithm based on a proposal distribution from linear Bayes theory. Section \ref{subsec:Software-trouble-reports} presents an industrial application to online prediction of faults in a large-scale software project, where allowing parameters to evolve
over time considerably improves predictive performance. Section \ref{subsec:Simulations} explores the properties of the inference method on simulated data. The final section concludes.

\section{Dynamic mixture of experts\label{sec:The model}}

Let $D_{j}=(y_{j},\mathbf{\tilde{x}}_{j})$ represent data from a
time dependent process observed at different time points $j=1,\ldots,J$,
where $y_{j}$ denotes the univariate response variable and $\mathbf{\tilde{x}}_{j}=(\tilde{x}_{j}^{(1)},\ldots,\tilde{x}_{j}^{(P)})^{\top}$
is a $P$-dimensional covariate vector. The $D_{j}$ may contain
only one observation as in standard time series applications, or it
may be a data batch containing several observations as in the software
upgrade process described in Section \ref{subsec:Software-trouble-reports}.
We propose the dynamic mixture of experts model 
\begin{equation}
f_{j}\left(y_{j}|\mathbf{\tilde{x}}_{j},\boldsymbol{\omega}_{j},\boldsymbol{\lambda}_{j}\right)=\sum_{k=1}^{K}\omega_{jk}\left(\mathbf{z}_{j}\right)f_{jk}\left(y_{j}|\lambda_{jk}\left(\mathbf{x}_{j}\right)\right),\label{eq:Dynamic model}
\end{equation}
for online (real time) prediction of $y_{j}$ given the value of the
covariate $\tilde{\boldsymbol{x}}_{j}$; $\mathbf{z}_{j}$ and $\mathbf{x}_{j}$
are subsets of $\mathbf{\tilde{x}}_{j}$ of dimensions $Q$ and $P$
respectively. The $\lambda_{jk}(\mathbf{x}_{j})$ and $\omega_{jk}(\mathbf{z}_{j})$,
$k=1,\ldots,K$, are time-varying covariate-dependent parameter and
mixture weight functions of the $k^{th}$ expert model respectively,
$\boldsymbol{\lambda}_{j}=(\lambda_{j1}(\mathbf{x}_{j}),\ldots,\lambda_{jK}(\mathbf{x}_{j}))$
and $\boldsymbol{\omega}_{j}=(\omega_{j2}(\mathbf{z}_{j}),\ldots,\omega_{jK}(\mathbf{z}_{j}))$.
The covariates in the mixture weights can be distinct from the covariates
in the experts.

Here, the experts represent the component models in the mixture \eqref{eq:Dynamic model} and they depend on the structure of the response variable; they are typically density functions from the exponential family, e.g.
Gaussian if $y_{j}$ is continuous, or Poisson, binomial or negative
binomial for count data, or multinomial if $y_{j}$ is categorical.
However, as in \citet{villani2012generalized}, we allow the component
models to be any well-behaved density functions, not necessarily limited
to the exponential family, and the model parameter may be multidimensional
with each of its components connected to the covariates through its
own link function. By well-behaved densities we mean densities that are twice differentiable with respect to the parameters, and that satisfies the nondegeneracy condition in \citet[Condition 1]{jiang1999identifiability} so that the mixture of experts model is identified. \citet {jiang1999identifiability} show that this condition is fulfilled for Poisson components and Section \ref{sec:identifiability} in the Appendix give a similar result for generalized Poisson components, which do not belong to the exponential family and are used in the empirical application in Section \ref{subsec:Software-trouble-reports}.

The component model parameters $\lambda_{jk}=\lambda_{jk}(\mathbf{x}_{j})$,
$k=1,\ldots,K$ are connected to their linear predictors through a
link function $g$ as

\begin{equation}
\eta_{jk}=g\left(\lambda_{jk}\right)=\mathbf{x}_{j}^{\top}\boldsymbol{\beta}_{jk},\label{eq:Mixture expected value}
\end{equation}
where $\mathbf{x}_{j}=(1,x_{j}^{(1)},...,x_{j}^{(P)})^{\top}$ and $\boldsymbol{\beta}_{jk}=(\beta_{jk}^{(0)},\ldots,\beta_{jk}^{(P)})^{\top}$.
For component models with more than one parameter, Eq. \ref{eq:Dynamic model}
can be extended by linking each parameter to its own linear predictor;
see \citet{villani2012generalized}. Furthermore, the mixture weights
depend on the covariate $\boldsymbol{z}_{j}$, through the multinomial
logit link function
\begin{equation}
\omega_{jk}=\frac{\exp\left(\psi_{jk}\right)}{1+\sum_{j=2}^{J}\exp\left(\psi_{jk}\right)},\label{eq: mixture weights link function}
\end{equation}
with 
\begin{equation}
\psi_{jk}=\mathbf{z}_{j}^{\top}\boldsymbol{\theta}_{jk},\,\,\,k=2,\ldots,K\label{eq:mixing weight predictor}
\end{equation}
where $\mathbf{z}_{j}=(1,z_{j}^{(1)},...,z_{j}^{(Q)})^{\top}$, and $\boldsymbol{\theta}_{jk}=(\theta_{jk}^{(0)},\ldots,\theta_{jk}^{(Q)})^{\top}$. Following standard practice we set $\psi_{j1}=0$ for all $j$ in \eqref{eq: mixture weights link function} to identify the model \citep[Remark 1]{jiang1999identifiability}.
In the following, we refer to $\boldsymbol{\beta}_{jk}$ and $\boldsymbol{\theta}_{jk}$
as the regression \emph{coefficients} in the component distributions
and mixture weights, respectively, and to $\boldsymbol{\lambda}_{j}=(\lambda_{j,1},\ldots,\lambda_{j,K})$
and $\boldsymbol{\omega}_{j}=(\omega_{j2},\ldots,\omega_{jK})$ as
the model \emph{parameters}. 

Mixture models are well-known to suffer from label switching, i.e. invariance under permutations of the components. A common approach is to impose order restrictions on parameters of the mixture components \citet{jiang1999identifiability}, either before running MCMC or SMC to sample from the posterior, or by re-ordering the posterior draws after the sampling \citep{stephens2000dealing}. Alternatively, in the Bayesian framework, the identifiability problem is addressed by designing informative priors \citep{malsiner2017identifying}. However, our interest here is on predictive inference, where label switching is not a concern \citep{geweke2007interpretation}.

\subsection{Prior process\label{subsec:Prior-process}}

To simplify notation, we stack all the
regression coefficients at time $j$ into one vector $\boldsymbol{\gamma}_{j}=(\boldsymbol{\beta}_{j}^{\top},\boldsymbol{\theta}_{j}^{\top})^{\top}$,
where $\boldsymbol{\beta}_{j}=(\boldsymbol{\beta}_{j1}^{\top},\ldots,\boldsymbol{\beta}_{jK}^{\top})$
and $\boldsymbol{\theta}_{j}=(\boldsymbol{\theta}_{j2}^{\top},\ldots,\boldsymbol{\theta}_{jK}^{\top})$,
and the linear predictors for all components into $\boldsymbol{\rho}_{j}=(\boldsymbol{\eta}_{j}^{\top},\boldsymbol{\psi}_{j}^{\top})^{\top}$. The prior for the $\boldsymbol{\gamma}_{j}$ is a random walk 

\begin{equation}
\boldsymbol{\gamma}_{j}=\boldsymbol{\gamma}_{j-1}+\boldsymbol{\varepsilon}_{j},\,\,\,\,\,\,\,\,\,\,\,\boldsymbol{\varepsilon}_{j}\sim N\left(0,\,\mathbf{U}_{j}\right),\label{eq:Prior distribution}
\end{equation}
with a predefined initial distribution $p(\boldsymbol{\gamma}_{1})$, which allows it to vary over time.
This prior process is commonly applied in dynamic models as a way of penalizing
the regression coefficients from high fluctuations and  avoiding overfitting \citep{fahrmeir2011bayesian,fahrmeir2004penalized}.
In some applications it is sufficient to set $\boldsymbol{U}_{j}=\boldsymbol{U}$,
which is a special case of (\ref{eq:Prior distribution}).
However, it is more useful for online inference to let $\boldsymbol{U}_{j}$
change over time as it allows to update the prior with historic data recursively as more data batches are observed. 

Fully Bayesian inference requires a prior for each $\boldsymbol{U}_{j}$.
Common priors are: i) an inverse-Wishart density for a full matrix
$\boldsymbol{U}_{j}$ \citep{gamerman1998markov}, ii) an inverse-gamma
density \citep{fahrmeir2004penalized} or a random walk process \citep{lang2002function}
on the elements of a diagonal $\boldsymbol{U}_{j}$. An alternative
to placing a prior on each $\boldsymbol{U}_{j}$ is to approximate
each $\boldsymbol{U}_{j}$ recursively using the discount factor approach
in \citet{west1985dynamic}. Let $\boldsymbol{C}_{j}$ denote the
posterior covariance of $\boldsymbol{\gamma}_{j}$ and set $\boldsymbol{U}_{j}=(\alpha^{-1}-1)\boldsymbol{C}_{j-1}$
for a given discount factor $0<\alpha<1$. A value of $\alpha$ close
to one shrinks $\boldsymbol{U}_{j}$ towards zero, leading to very
little variation in $\boldsymbol{\gamma}_{j}$ over time; a value
of $\alpha$ close to zero gives the regression parameters more flexibility
and allows the model to adapt well to local fluctuations in the parameter;
for instance, change points or level shifts in the parameter. 

The discount factor approach has some advantages compared to a fully
Bayesian approach. It is computationally much quicker as it avoids
extra simulations from the posterior of $\boldsymbol{U}_{j}$. The
discount factor conveniently controls the smoothness of the parameter
evolution through time with a single parameter, and it allows building
static and dynamic models in a unified way just by changing the value
of $\alpha$. Following \citet{liu2001combined}, models with $.95\leq\alpha<1$
are essentially static, and those with $\alpha<.95$ are dynamic.
We use this approach in Sections  \ref{subsec:Software-trouble-reports} and \ref{subsec:Simulations}. 

Our inference methodology applies also to the case of a fully Bayesian
approach where $\boldsymbol{U}_{j}$ is estimated in an additional
step using particle Markov chain Monte
Carlo \citep{andrieu2010particle} and SMC2 \citep{chopin2013smc2} 
methods which allow inference in models with both fixed and time-varying
(latent) parameters. It can also be used in the online parameter learning methodology of \citet{carvalho2010particle}.

\section{Inference, prediction and model comparison\label{sec:Inference}}

The state space model in Section \ref{sec:The model} enables us to exploit the vast literature \citep{gordon1993novel,pitt1999filtering,doucet2000sequential,doucet2006efficient,doucet2009tutorial,klaas2012toward}
available on sequential Monte Carlo (SMC). SMC methods are particularly
appropriate for sampling from the online posterior and real-time predictive
distributions. The present model often has many parameters,
and off-the-shelf SMC algorithms with simple proposal distributions
like the bootstrap filter \citep{gordon1993novel} will therefore perform poorly. This section
describes our proposed algorithm for sampling from the online posterior
using a particle filter tailored specifically to the class of dynamic
mixture of experts models. We also present the model comparison criteria
used to select the number of mixture components and the discount factor.

\subsection{The marginal particle filter approximation of the online posterior
distribution}

The target density is the online posterior $p\left(\boldsymbol{\gamma}_{j}|\,D_{1:j}\right)$  updated sequentially  in time using a \emph{prediction step}

\begin{equation}
p\left(\boldsymbol{\gamma}_{j}|\,D_{1:j-1}\right)=\int p\left(\boldsymbol{\gamma}_{j}|\,\boldsymbol{\gamma}_{j-1}\right)p\left(\boldsymbol{\gamma}_{j-1}|\,D_{1:j-1}\right)d\boldsymbol{\gamma}_{j-1},\label{eq:Prior predictive-1}
\end{equation}
followed by a \emph{measurement update step} using Bayes' theorem 
\begin{align}
p\left(\boldsymbol{\gamma}_{j}|\,D_{1:j}\right) & \propto f_{j}\left(y_{j}|\mathbf{\tilde{x}}_{j},\boldsymbol{\gamma}_{j}\right)p\left(\boldsymbol{\gamma}_{j}|\,D_{1:j-1}\right),\label{eq:Filtering distribution-1}
\end{align}
to make prior-to-posterior updates. The function $f_{j}(\cdot)$ is the response density defined in \eqref{eq:Dynamic model}, $D_{1:j}$ denotes the data observed until time $j$, and $p\left(\boldsymbol{\gamma}_{j}|\,D_{1:j-1}\right)$ is the prior updated with all historic data observed before the data batch $D_{j}$. Note that contrary to  \eqref{eq:Dynamic model}, now $f_{j}(\cdot)$ is parametrized in terms of only the regression coefficients $\boldsymbol{\gamma}_{j}$; this is because all quantities required in the inference methodology discussed later are expressed in terms of $\boldsymbol{\gamma}_{j}$ only.

 We are interested in the online predictive distribution $p\left(y_{j}|\mathbf{\tilde{x}}_{j},\boldsymbol{y}_{1:j-1}\right)$
which only depends on the filtering density up to time $j-1$ \citep{doucet2000sequential}. However, the challenging part of the sequential inference in (\ref{eq:Prior predictive-1}) - (\ref{eq:Filtering distribution-1}) is that
the integral in (\ref{eq:Prior predictive-1}) is only tractable for linear Gaussian models \citep{west1985dynamic,gordon1993novel}. 
One way to sample from (\ref{eq:Filtering distribution-1}) is to use a particle  filter algorithm. The particle filter is very attractive for real-time predictions; it allows to sample from intractable distributions and it does not require a scan of the full dataset every time a new observations becomes available.

We use the marginal particle filter of \citet{klaas2012toward}  to generate a set of  particles  $\{\boldsymbol{\gamma}_{j}^{m}\}_{m=1}^{M}$ associated with the importance weights $\{w_{j}^{m}\}_{m=1}^{M}$. Given the particle sample, any posterior expectation 
\begin{equation}
E\left(h(\gamma_{j})\right)=\int h(\gamma_{j})p(\gamma_{j}\vert D_{1:j})d\gamma_{j}
\end{equation}
is approximated sequentially by
\begin{equation}
\widehat{E}\left(h(\gamma_{j})\right)=\frac{\sum_{m=1}^{M}h(\gamma_{j}^{m})w_{j}^{m}}{\sum_{m=1}^{M}w_{j}^{m}}. \label{eq:IS_estimator}
\end{equation}
 The estimator \eqref{eq:IS_estimator} converges to  $E\left(h(\gamma_{j})\right)$ as $M\rightarrow\infty$ under some weak assumptions stated in \citet{geweke1989bayesian}; see \citet{doucet2001sequential}, and  \citet{chopin2004central} for more results on the convergence of the particle filter. 

The marginal particle filter proposes particles from the proposal distribution $q(\boldsymbol{\gamma}_{j}|\,D_{1:j})$ and computes the importance weights as

\begin{equation}
w_{j}^{m}\propto\frac{f_{j}\left(y_{j}|\mathbf{\tilde{x}}_{j},\boldsymbol{\gamma}_{j}^{m}\right)\sum_{h=1}^{M}w_{j-1}^{h}p\left(\boldsymbol{\gamma}_{j}^{m}|\,\boldsymbol{\gamma}_{j-1}^{h}\right)}{q\left(\boldsymbol{\gamma}_{j}^{m}|\,D_{1:j}\right)}.\label{Importance weights}
\end{equation}
The estimation of the density in \eqref{eq:Prior predictive-1} follows Eq.\eqref{eq:IS_estimator}; i.e,
\begin{equation}
\widehat{p}\left(\boldsymbol{\gamma}_{j}|\,D_{1:j-1}\right)=\frac{\sum_{m=1}^{M}w_{j-1}^{m}p\left(\boldsymbol{\gamma}_{j}|\,\boldsymbol{\gamma}_{j-1}^{m}\right)}{\sum_{m=1}^{M}w_{j}^{m}}\label{approximate prior predictive}
\end{equation}

Notice that the importance weights in \eqref{Importance weights} depend on the likelihood $f_{j}(y_{j}|\mathbf{\tilde{x}}_{j},\boldsymbol{\gamma}_{j})$
expressed using the covariates and the regression coefficients rather than the model parameters as in (\ref{eq:Dynamic model}); the inference requires keeping track of the regression coefficients only.

Clearly the proposal density plays an important role. A proposal density which is inconsistent with the target posterior may lead to the particle degeneracy: The importance weights of only a few particles tend to be substantially different from zero, leading to very few effective samples.  To mitigate this degeneracy issue, particles with low weights are discarded and replaced by copies of the particles with high weights. Various strategies for resampling  particles are available in the literature \citep{gordon1993novel,liu1998sequential,carpenter1999improved,fearnhead2003line} and \citet{douc2005comparison} compare some of these resampling schemes. 

In the next section we use linear Bayes methods \citep{west1985dynamic}
to construct a proposal  $q(\boldsymbol{\gamma}_{j}|\,D_{1:j})$  that is tailored to the true posterior, which
is crucial for particle methods in high-dimensional parameter spaces.

\subsection{A computationally fast proposal distribution for high-dimensional
marginal particle filters\label{subsec:Proposal-distribution-based-1}}

\citet{west1985dynamic} develop a linear Bayes method \citep{goldstein2007bayes}
for dynamic generalized linear models with recursions for the posterior
mean and covariance over time, making no assumptions on the distributional
form of the posterior. \citet{ravines2007efficient} use these recursive
moments to design a multi-move proposal for MCMC targeting the joint
smoothing posterior in dynamic generalized linear models. We combine
the linear Bayes method in \citet{west1985dynamic} with ideas from
the EM algorithm \citet[ch.9]{bishop2006pattern} to design a proposal
distribution $q\left(\boldsymbol{\gamma}_{j}|D_{1:j}\right)$ targeting
the filtering density $p(\boldsymbol{\gamma}_{j}|\,D_{1:j})$ in dynamic
mixture of experts models. The proposed method allows general mixture
components outside the exponential family with any twice differentiable
link function.

Similar to Eq. (2.8) in \citet{west1985dynamic}, we can write the joint
posterior of the regression coefficients and the linear predictors
as
\begin{equation}
p(\boldsymbol{\gamma}_{j},\boldsymbol{\rho}_{j}\vert D_{1:j})=p(\boldsymbol{\rho}_{j}\vert D_{1:j})p(\boldsymbol{\gamma}_{j}\vert\boldsymbol{\rho}_{j},D_{1:j-1}),\label{eq:posteriorBetaTheta}
\end{equation}
where we recall that $\boldsymbol{\rho}_{j}=(\boldsymbol{\eta}_{j}^{\top},\boldsymbol{\psi}_{j}^{\top})^{\top}$contains
the linear predictors in all components and mixture weights. The second
factor in (\ref{eq:posteriorBetaTheta}) does not condition on $D_{j}$
since $\boldsymbol{\gamma}_{j}$ only enters the likelihood function
through the scalar-valued linear predictors in each component, $\eta_{jk}=\mathbf{x}_{j}^{\top}\boldsymbol{\beta}_{jk}$
and $\psi_{jk}=\mathbf{z}_{j}^{\top}\boldsymbol{\theta}_{jk}$ for $k=1,\ldots,K.$
Our proposal is tailored to the posterior $p\left(\boldsymbol{\gamma}_{j}|D_{1:j}\right)$
by using the following steps:

\begin{enumerate}
\item Approximate the prior $p(\boldsymbol{\gamma}_{j}\vert D_{1:j-1})$
using a Gaussian with mean and covariance computed from particles
at time $j-1$. 
\item Obtain the second factor in (\ref{eq:posteriorBetaTheta}) by conditioning
$p(\boldsymbol{\gamma}_{j}\vert D_{1:j-1})$ on the linear restrictions
$\boldsymbol{\rho}_{j}$.
\item Propose from $q(\boldsymbol{\gamma}_{j}|D_{1:j})=N(\boldsymbol{\mu}_{j},\boldsymbol{H}_{j})$,
where $\boldsymbol{\mu}_{j}$ and $\boldsymbol{H}_{j}$ are obtained
from the law of iterated expectation and law of total variance on
(\ref{eq:posteriorBetaTheta}) using a Gaussian approximation of $p(\boldsymbol{\rho}_{j}\vert D_{1:j})$. 
\end{enumerate}

To give the details of the three steps, define $\boldsymbol{\eta}_{j}:=\boldsymbol{X}_{j}\boldsymbol{\beta}_{j}$,
where $\boldsymbol{X}_{j}=I_{K}\otimes\boldsymbol{x}_{j}^{\top}$ and
$\boldsymbol{\psi}_{j}:=\boldsymbol{Z}_{j}\boldsymbol{\theta}_{j}$,
where $\boldsymbol{Z}_{j}=I_{K}\otimes\boldsymbol{z}_{j}^{\top}$; hence, 
 we can compactly write $\boldsymbol{\rho}_{j}=\boldsymbol{W}_{j}\boldsymbol{\gamma}_{j}$
where $\boldsymbol{\gamma}_{j}=(\boldsymbol{\beta}_{j}^{\top},\boldsymbol{\theta}_{j}^{\top})^{\top}$,
$\boldsymbol{\rho}_{j}=(\boldsymbol{\eta}_{j}^{\top},\boldsymbol{\psi}_{j}^{\top})^{\top}$
and 
\[
\boldsymbol{W}_{j}=\left(\begin{array}{cc}
I_{K}\otimes\boldsymbol{x}_{j}^{\top} & \boldsymbol{0}\\
\boldsymbol{0} & I_{K}\otimes\boldsymbol{z}_{j}^{\top}
\end{array}\right).
\]

We can use particles from time step $j-1$ to approximate $\boldsymbol{\gamma}_{j}|D_{1:j-1}\sim N(\boldsymbol{\bar{\gamma}}_{j},\Sigma_{\boldsymbol{\gamma}_{j}})$,
where 
\begin{equation}
\boldsymbol{\bar{\gamma}}_{j}=\sum_{m=1}^{M}w_{j-1}^{m}\boldsymbol{\gamma}_{j-1}^{m},\,\,\,\,\,\Sigma_{\boldsymbol{\gamma}_{j}}=\mathbf{U}_{j}+\sum_{m=1}^{M}w_{j-1}^{m}\left(\boldsymbol{\gamma}_{j-1}^{m}-\boldsymbol{\bar{\gamma}}_{j}\right)\left(\boldsymbol{\gamma}_{j-1}^{m}-\boldsymbol{\bar{\gamma}}_{j}\right)^{2},\label{eq: prior moments}
\end{equation}
 and then obtain the mean and covariance of the second factor of (\ref{eq:posteriorBetaTheta})
by conditioning this distribution on the linear constraints $\boldsymbol{\rho}_{j}=\boldsymbol{W}_{j}\boldsymbol{\gamma}_{j}$
\citep[eq. 2.28-2.29]{rue2005gaussian} yielding
\begin{align*}
E\left[\boldsymbol{\gamma}_{j}\vert\boldsymbol{\rho}_{j},D_{1:j-1}\right] & =\bar{\boldsymbol{\boldsymbol{\gamma}}}_{j}+\Sigma_{\boldsymbol{\gamma}_{j}\rho_{j}}\Sigma_{\rho_{j}}^{-1}(\boldsymbol{\rho}_{j}-\bar{\boldsymbol{\rho}}_{j})\\
V\left[\boldsymbol{\gamma}_{j}\vert\boldsymbol{\rho}_{j},D_{1:j-1}\right] & =\Sigma_{\boldsymbol{\gamma}_{j}}-\Sigma_{\boldsymbol{\gamma}_{j}\rho_{j}}\Sigma_{\rho_{j}}^{-1}\Sigma_{\rho_{j}\boldsymbol{\gamma}_{j}},
\end{align*}
where $\bar{\boldsymbol{\rho}}_{j}=\boldsymbol{W}_{j}\bar{\boldsymbol{\gamma}}_{j}$,
$\Sigma_{\rho_{j}}=\boldsymbol{W}_{j}\Sigma_{\boldsymbol{\gamma}_{j}}\boldsymbol{W}_{j}^{\top}$,
$\Sigma_{\rho_{j}\boldsymbol{\gamma}_{j}}=\boldsymbol{W}_{j}\Sigma_{\boldsymbol{\gamma}_{j}}$,
$\Sigma_{\boldsymbol{\gamma}_{j}\rho_{j}}=\Sigma_{\boldsymbol{\gamma}_{j}}\boldsymbol{W}_{j}^{\top}$.
Now, the proposal is $q\left(\boldsymbol{\gamma}_{j}|D_{1:j}\right)=N\left(\boldsymbol{\mu}_{j},\boldsymbol{H}_{j}\right)$
with moments obtained from applying the law of iterated expectations
and the law of total variance to (\ref{eq:posteriorBetaTheta}),

\begin{align}
\boldsymbol{\mu}_{j} & =E_{\boldsymbol{\rho}_{j}}\left[E\left[\boldsymbol{\gamma}_{j}\vert\boldsymbol{\rho}_{j},D_{1:j-1}\right]\vert D_{1:j}\right]=\bar{\boldsymbol{\boldsymbol{\gamma}}}_{j}+\Sigma_{\boldsymbol{\gamma}_{j}\rho_{j}}\Sigma_{\rho_{j}}^{-1}\left(E_{\boldsymbol{\rho}_{j}}(\boldsymbol{\rho}_{j}\vert D_{1:j})-\bar{\boldsymbol{\rho}}_{j}\right)\label{eq: Posterior mean of reg coef}\\
\boldsymbol{H}_{j} & =E_{\boldsymbol{\rho}_{j}}\left[V\left[\boldsymbol{\gamma}_{j}\vert\boldsymbol{\rho}_{j},D_{1:j-1}\right]\vert D_{1:j}\right]+V_{\boldsymbol{\rho}_{j}}\left[E\left[\boldsymbol{\gamma}_{j}\vert\boldsymbol{\rho}_{j},D_{1:j-1}\right]\vert D_{1:j}\right]\label{eq: Posterior var of reg coef}\\
 & =\Sigma_{\boldsymbol{\gamma}_{j}}-\Sigma_{\boldsymbol{\gamma}_{j}\rho_{j}}\left(\Sigma_{\rho_{j}}^{-1}-\Sigma_{\rho_{j}}^{-1}V_{\boldsymbol{\rho}_{j}}(\boldsymbol{\rho}_{j}\vert D_{1:j})\Sigma_{\rho_{j}}^{-1}\right)\Sigma_{\rho_{j}\boldsymbol{\gamma}_{j}}\nonumber 
\end{align}
It remains to compute $E_{\boldsymbol{\rho}_{j}}(\boldsymbol{\rho}_{j}\vert D_{1:j})$
and $V_{\boldsymbol{\rho}_{j}}(\boldsymbol{\rho}_{j}\vert D_{1:j})$.
A second order Taylor expansion of $\log p(\boldsymbol{\rho}_{j}\vert D_{1:j})$
around $\bar{\boldsymbol{\rho}}_{j}$ leads to the following approximations
\citep{doucet2000sequential}: 
\[
V_{\boldsymbol{\rho}_{j}}(\boldsymbol{\rho}_{j}\vert D_{1:j})=\left[-\left.\nabla_{}\nabla_{\boldsymbol{\rho}_{j}}\log p(\boldsymbol{\rho}_{j}\vert D_{1:j})\right|_{\boldsymbol{\rho}_{j}=\bar{\boldsymbol{\rho}}_{j}}\right]^{-1},
\]
\begin{equation}
E_{\boldsymbol{\rho}_{j}}(\boldsymbol{\rho}_{j}\vert D_{1:j})=\bar{\boldsymbol{\rho}}_{j}+V_{\boldsymbol{\rho}_{j}}(\boldsymbol{\rho}_{j}\vert D_{1:j})\left.\nabla_{\boldsymbol{\rho}_{j}}\log p(\boldsymbol{\rho}_{j}\vert D_{1:j})\right|_{\boldsymbol{\rho}_{j}=\bar{\boldsymbol{\rho}}_{j}}.\label{eq:linear predictor estimator}
\end{equation}
Letting $\pi_{jk}=\log\omega_{jk}f_{jk}\left(y_{j}|\lambda_{jk}\right)$,
the gradient can be computed by direct calculation 
\begin{align*}
\nabla_{\boldsymbol{\rho}_{j}}\log p(\boldsymbol{\rho}_{j}\vert D_{1:j}) & =\sum_{k=1}^{K}\mathrm{Pr}(s_{j}=k\vert D_{1:j})\nabla_{\boldsymbol{\rho}_{j}}\pi_{jk}-\Sigma_{\rho_{j}}^{-1}(\boldsymbol{\rho}_{j}-\bar{\boldsymbol{\rho}}_{j}),
\end{align*}
where $\mathrm{Pr}(s_{j}=k\vert D_{1:j})\propto\omega_{jk}f_{jk}(y_{j}\vert D_{1:j-1},\lambda_{jk})$
are the posterior probabilities of the observation $y_{j}$ coming
from component $k$ (see \citealp[ch. 9.3]{bishop2006pattern} for
similar expressions for the EM algorithm). Similarly, the Hessian
is, 
\begin{align*}
\nabla\nabla_{\boldsymbol{\rho}_{j}}\log p(\boldsymbol{\rho}_{j}\vert D_{1:j}) & =\sum_{k=1}^{K}\mathrm{Pr}(s_{j}=k\vert D_{1:j})\nabla\nabla_{\boldsymbol{\rho}_{j}}\pi_{jk}-\Sigma_{\rho_{j}}^{-1}
\end{align*}
Note that the component parameters $\eta_{jk}$ and $\psi_{jk}$ enter
additively in $\log\omega_{jk}f_{jk}\left(y_{j}|\lambda_{jk}\right)$;
therefore, their gradients can be computed separately. 

If the batches $D_{j}$ contain several observations, then $\boldsymbol{\mu}_{j}$
and $\boldsymbol{H}_{j}$ can be computed by iterating the procedure
described above over the observations in the batch; see \citet{gamerman1991dynamic}
for a similar approach. Starting with the first observation, we proceed
through the following iterations:
\begin{enumerate}
\item Compute $\boldsymbol{\mu}_{j}^{(i)}$ and $\boldsymbol{H}_{j}^{(i)}$
from Eq. \ref{eq: Posterior mean of reg coef} and Eq. \ref{eq: Posterior var of reg coef}.
\item Set $\bar{\boldsymbol{\gamma}}_{j}=\boldsymbol{\mu}_{j}^{(i)}$ and
$\Sigma_{\boldsymbol{\gamma}_{j}}=\boldsymbol{H}_{j}^{(i)}$.
\item Return to step $1$ until the last observation in the batch. 
\end{enumerate}

\subsection{Model comparison and prediction \label{subsec:Model-comparison-and}}

Our model depends on the choice of the number of mixture components
$K$ and the discount factor $\alpha$. We propose to infer those
quantities using a sequential version of the marginal likelihood \citep{doucet2000sequential}
\newpage
\begin{equation}
p\left(y_{1:J}\right)=p\left(y_{1}\right)\prod_{j=2}^{J}p\left(y_{j}|y_{1:j-1}\right),\label{eq:sequential marginal likelihood-1}
\end{equation}
 where
\begin{align}
p\left(y_{j}|y_{1:j-1}\right) & =\int f_{j}\left(y_{j}|\mathbf{\tilde{x}}_{j},\boldsymbol{\gamma}_{j}\right)p\left(\boldsymbol{\gamma}_{j}|\,D_{1:j-1}\right)d\boldsymbol{\gamma}_{j}.\label{eq:likelihood estimate-1}
\end{align}
Given a sample of $M$ particles $\{\boldsymbol{\gamma}_{j-1}^{m}\}_{m=1}^{M}$
and the corresponding importance weights $\{w_{j-1}^{n}\}_{n=1}^{N}$,
the predictive distribution (\ref{eq:likelihood estimate-1}) is approximated
as
\[
\hat{p}\left(\boldsymbol{y}_{j}|\boldsymbol{y}_{1:j-1}\right)=\sum_{m=1}^{M}w_{j-1}^{m}f_{j}\left(y_{j}|\mathbf{\tilde{x}}_{j},\gamma_{j}^{m}\right),
\]
where $\gamma_{j}^{m}$ are generated from the transition distribution
$p(\boldsymbol{\gamma}_{j}|\boldsymbol{\gamma}_{j-1}^{m})$.
Different predictive scores are defined as functions of \eqref{eq:sequential marginal likelihood-1}. One particular example is the log predictive score 
\[
LPS=\sum_{j=J^{*}}^{J}\log\hat{p}\left(\boldsymbol{y}_{j}|\boldsymbol{y}_{1:j-1}\right),
\] where $J^{*} \in [1,\ldots,J)$. The LPS is generally sensitive to the initial distribution of the parameters \citep{villani2009regression}. We therefore use the last  $\nicefrac{J}{2}$  data batches to compute the LPS for the models in Section \ref{subsec:Software-trouble-reports}; i.e $J^{\star} = \nicefrac{J}{2}$. We assume that the particle approximation to the marginal likelihood should be stable after $j=\nicefrac{J}{2}$. Computing the $LPS$ for different combinations of the number of mixture components $K$ and the discount factor $\alpha$ makes it possible to select good values for these model specification parameters.

\section{Predicting faults in large-scale software projects\label{subsec:Software-trouble-reports}}

Large-scale industrial software projects are continually upgraded to fix bugs and/or to add new features. The upgrades are generally at irregular times: in one week we may observe one release, and in the next, two or three releases depending on several factors such as the amount and severity of the bugs reported in previous versions, the complexity of the new features added to the software, and other business-related factors. Other key factors include the human interaction with the software and the technology evolution. The developers, the user behavior and technologies change over time. The dynamic mixture of experts model (\ref{eq:Dynamic model}) is appropriate in this case. Different mixture components allow us to model the unknown variations/changes in the human interaction with the software and the time-varying parameters enable the model to adapt to the changes over time.

As the response variable is the number of faults $y_{t}$ reported on the upgrade created at time $t$, we propose a dynamic mixture of Poisson experts. Here each expert is a Poisson regression model with a covariate vector $\mathbf{\tilde{x}}_{t}$ selected from six code complexity metrics that measure changes made in the source code. The metrics include: i) The number of commits (NC) which represents the number of modifications done from the previous to the current version, ii) the number of changed modules (CM), iii) the number of faults corrected (NFC) per line of code which is the ratio of the total number of faults corrected and the total number of code lines excluding comments, iv) the proportion of files written in C++ (CF), v) the proportion of files written in Java (JF), and vi) the file complexity (FC). The latter is a score calculated based on the number of control flows in the code, e.g. if, for and while loop statements.
 
The aim is to build an online prediction model for the number of faults in a planned upgrade release. We use a software trouble reports data set
from a large-scale project at a major telecom company; the dataset contains a history of $1800$ upgrades that were created during a
period of $650$ days (roughly $21$ months). All covariates, excluding the CF, JF and NFC are integers ranging from zero to a value up to six order of magnitude. Therefore, to reduce the scale variations, we apply the $\log(1+\tilde{x}_{t})$ transformation to the integer complexity metrics; after this transformation the highest value is no greater than $15$. 

\begin{figure}[H]
\subfloat[]{\includegraphics[scale=0.8]{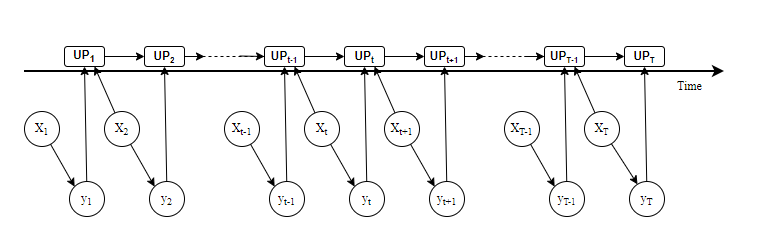}}

\subfloat[]{\includegraphics[scale=0.8]{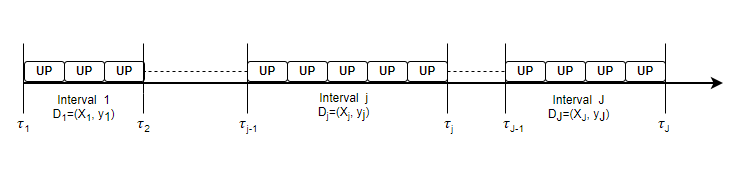}

}

\caption{{\small{}(a) The upgrading process. An upgrade (UP) at time $t$ is
created by making $x_{t}$ changes on the previous version of the
software (created at time $t-1$) and $y_{t}$ faults are reported
on the version created at time $t$. (b) Process of grouping upgrades
according to intervals partitioning the training time.}\label{fig:Upgrade-packages-grouped}}
\end{figure}

To make it tractable to deal with the  irregular times of fault reports, we partition time into short contiguous intervals $[\tau_{0},\tau_{1}),[\tau_{1},\tau_{2})\ldots,[\tau_{J-1},\tau_{J})$, where $\tau_{0}=\min(t)<\tau_{1}<\ldots<\tau_{J-1}<\tau_{J}=\max(t)$. The partition of time induces a partition of the original data into a sequence of batches  $D_{j}=\left\{ \mathbf{y}_{j},\mathbf{\tilde{X}}_{j}\right\} $  which collect data for all upgrade packages created within the time
interval $t\in[\tau_{j-1},\tau_{j}),\,j=1\ldots,J$. Batch  $D_{j}$ contains  $N_{j}$ data points, where $\boldsymbol{y}_{j}=(y_{1j},\ldots,y_{N_{j},j})^{\top}$ is a vector of the response observations in the batch and $\mathbf{\tilde{X}}_{j}=(\mathbf{\tilde{x}}_{1j},\ldots,\tilde{\mathbf{x}}_{N_{j},j})^{\top}$ is a vector of covariates $\mathbf{\tilde{x}}_{ij}$ for the data point $i = 1, . . . , N_{j}$. Figure \ref{fig:Upgrade-packages-grouped} illustrates this data partition.

The time is partitioned into $30$ days-long intervals, which leads to $21$ intervals in total. Experimentation with intervals lengths
of one week, two weeks and three months did not improve the LPS. We also assume the initial distribution $\boldsymbol{\gamma}_{1}\sim N(0,\,\,I)$, where $I$ is the unit diagonal matrix; see Appendix \ref{sec:Examples-of-component} for details.

Table \ref{tab:Log-predictive-score} compares different fitted models based on their LPS. The table displays various dynamic models, with
discount factor $\alpha=0.5$, and their static versions, where $\alpha=0.99$. The models in the table have different variables in the component
models and the number of commits (NC) as the only covariate $\mathbf{z}$ in the mixture weights. To select $\mathbf{z}$, we fix the covariates
in the component models to $\tilde{X}$ and, starting from $\mathbf{z}=\tilde{X}$, we eliminate variables in $\mathbf{z}$ systematically based on the LPS.

\begin{center}
\begin{table}[H]
{\footnotesize{}\caption{{\small{}LPS for different models fitted to the software trouble reports
data.}{\footnotesize{} }{\small{}Results are based on a posterior
sample of $100000$ particles}{\footnotesize{}.\label{tab:Log-predictive-score} }}
}{\footnotesize\par}
\centering{}{\footnotesize{}}%
\begin{tabular}{llccccc}
\hline 
\multirow{2}{*}{\textbf{\footnotesize{}Component model}} & \multirow{2}{*}{\textbf{\footnotesize{}Type}} &  & \multicolumn{3}{c}{\textbf{\footnotesize{}Number of components}} & \tabularnewline
\cline{3-7} 
 &  &  & \textbf{\footnotesize{}1} & \textbf{\footnotesize{}2} & \textbf{\footnotesize{}3} & \textbf{\footnotesize{}4}\tabularnewline
\hline 
\multirow{2}{*}{{\footnotesize{}CM}} & \textbf{\footnotesize{}Dynamic} &  & \textbf{\footnotesize{}$\boldsymbol{-1550.71}$} & \textbf{\footnotesize{}$\boldsymbol{-1192.66}$} & \textbf{\footnotesize{}$\boldsymbol{-1179.44}$} & {\footnotesize{}$\boldsymbol{-1173.35}$}\tabularnewline
\cline{2-7} 
 & {\footnotesize{}Static} &  & {\footnotesize{}$-1586.76$} & {\footnotesize{}$-1345.49$} & {\footnotesize{}$-1283.65$} & {\footnotesize{}$-1299.76$}\tabularnewline
\hline 
\multirow{2}{*}{{\footnotesize{}CM + FC}} & \textbf{\footnotesize{}Dynamic} &  & {\footnotesize{}$\boldsymbol{-1539.21}$} & {\footnotesize{}$\boldsymbol{-1168.96}$} & {\footnotesize{}$\boldsymbol{-1160.85}$} & {\footnotesize{}$\boldsymbol{-1160.73}$}\tabularnewline
\cline{2-7} 
 & {\footnotesize{}Static} &  & {\footnotesize{}$-1579.58$} & {\footnotesize{}$-1486.43$} & {\footnotesize{}$-1303.87$} & {\footnotesize{}$-1281.43$}\tabularnewline
\hline 
\multirow{2}{*}{{\footnotesize{}CM+FC+NC}} & \textbf{\footnotesize{}Dynamic} &  & {\footnotesize{}$\boldsymbol{-1543.58}$} & {\footnotesize{}$\boldsymbol{-1187.99}$} & {\footnotesize{}$\boldsymbol{-1170.84}$} & {\footnotesize{}$\boldsymbol{-1278.04}$}\tabularnewline
\cline{2-7} 
 & {\footnotesize{}Static} &  & {\footnotesize{}$-1688.55$} & {\footnotesize{}$-1279.10$} & {\footnotesize{}$-1269.48$} & {\footnotesize{}$-1307.74$}\tabularnewline
\hline 
\multirow{2}{*}{{\footnotesize{}CM+FC+NC+NFC}} & \textbf{\footnotesize{}Dynamic} &  & {\footnotesize{}$\boldsymbol{-1543.03}$} & {\footnotesize{}$\boldsymbol{-1212.97}$} & {\footnotesize{}$\boldsymbol{-1247.81}$} & {\footnotesize{}$\boldsymbol{-1290.10}$}\tabularnewline
\cline{2-7} 
 & {\footnotesize{}Static} &  & {\footnotesize{}$-1608.61$} & {\footnotesize{}$-1364.98$} & {\footnotesize{}$-1319.60$} & {\footnotesize{}$-1281.99$}\tabularnewline
\hline 
\multirow{2}{*}{{\footnotesize{}CM+FC+NC+NFC+JF}} & \textbf{\footnotesize{}Dynamic} &  & {\footnotesize{}$\boldsymbol{-1534.11}$} & {\footnotesize{}$\boldsymbol{-1250.82}$} & {\footnotesize{}$\boldsymbol{-1288.74}$} & {\footnotesize{}$\boldsymbol{-1272.16}$}\tabularnewline
\cline{2-7} 
 & {\footnotesize{}Static} &  & {\footnotesize{}$-1596.00$} & {\footnotesize{}$-1366.89$} & {\footnotesize{}$-1305.95$} & {\footnotesize{}$-1259.47$}\tabularnewline
\hline 
\multirow{2}{*}{{\footnotesize{}CM+FC+NC+NFC+JF+CF}} & \textbf{\footnotesize{}Dynamic} &  & {\footnotesize{}$\boldsymbol{-1537.74}$} & {\footnotesize{}$\boldsymbol{-1243.42}$} & {\footnotesize{}$\boldsymbol{-1315.38}$} & {\footnotesize{}$\boldsymbol{-1268.29}$}\tabularnewline
\cline{2-7} 
 & {\footnotesize{}Static} &  & {\footnotesize{}$-1616.99$} & {\footnotesize{}$-1370.08$} & {\footnotesize{}$-1326.20$} & {\footnotesize{}$-1304.69$}\tabularnewline
\hline 
\end{tabular}{\footnotesize\par}
\end{table}
\par\end{center}

Table \ref{tab:Log-predictive-score} shows that dynamic models outperform static models, with a difference in LPS more that $40$ for single component models and well above $100$ for several of the multicomponent models. Also, there is a very large jump in LPS when going from one to two components, in particular for the dynamic versions. While two components seem to be sufficient for the dynamic models, the static models require more components and covariates. The dynamic model CM+FC with two components seems to perform well in terms of LPS  since adding more complexity gives no significant increase in LPS. This model is therefore selected for further analysis. 

To illustrate that our algorithm can be used also for models outside the exponential family we fit a single component dynamic generalized Poisson model  \citep{famoye2006zero}  using CM and FC as covariates in both the mean and dispersion functions; see Appendix \ref{sec:generalized poisson details}. Figure \ref{fig:Predictive distribution} displays the predictive distribution for the one and two-component versions of the selected CM+FC dynamic model, and the dynamic generalized Poisson model at three time points: $j=2$, $j=10$ and $j=21$. The predictive distribution at the time point $j$ is constructed using the posterior at the previous interval $j-1$ and the batch $D_{j}$ as test set.

\begin{figure}[H]
\begin{singlespace}
\noindent \begin{centering}
\includegraphics[scale=0.85]{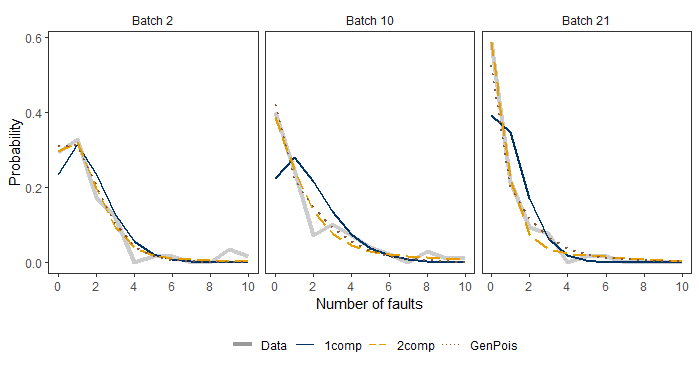}
\par\end{centering}
\end{singlespace}
\caption{{\small{}The evolution of the predictive distribution of one-component and two-components of the CM+FC dynamic Poisson ME model and the one-component dynamic generalized Poisson model fitted to the software fault data at three time points $j=2$, $j=10$, and $j=21$.}{\footnotesize{}\label{fig:Predictive distribution}}}
\end{figure}

It is clear from Figure \ref{fig:Predictive distribution} that the distribution of the number of faults varies over time; there is a very large shift of probability mass toward a smaller number of faults as time evolve. The two-component CM+FC dynamic model adapts well to the dynamic variations in the data and gives very impressive predictions on the test data, while the one-component version does not perform well, agreeing with the LPS in Table \ref{tab:Log-predictive-score}.  The one-component dynamic generalized Poisson model behaves very similarly to the two-component dynamic Poisson; the LPS of the generalized Poisson model is $-1189$.

\begin{figure}[H]
\begin{centering}
\includegraphics[scale=0.8]{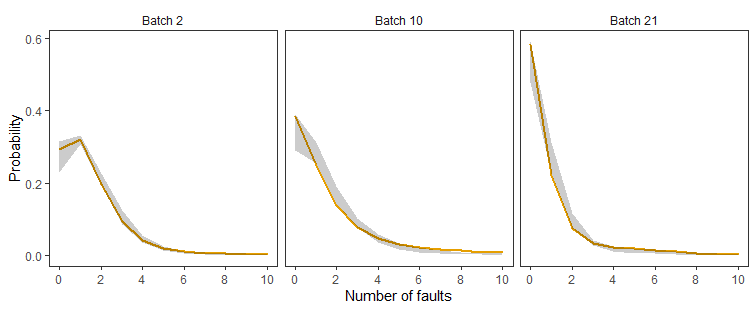}
\par\end{centering}
\caption{{\small{}Illustrating the efficiency of the proposed inference
methodology. The shaded area represents the $95\%$
intervals of the predictions from $100$ independent iterations of
the particle filter with $1000$ particles. The orange curve represents
the prediction obtained from a particle filter with $100,000$ particles.}{\footnotesize{}\label{fig:EvolutionPredDensity}}}
\end{figure}

To investigate the efficiency of the proposed SMC inference methodology, we fit the selected two-component  CM+FC  dynamic model using a particle filter with $1000$ particles. This is at least an order of magnitude smaller than what can easily be afforded in real applications, but is used here to investigate how much the inferred predictive distribution varies over $100$ independent runs with different seeds. 
Figure \ref{fig:EvolutionPredDensity} shows that this variability is small; the figure
also includes the predictive distribution from a single run with $100,000$ particles to represent the ground truth. This shows that the proposed method is very efficient and even a small number of particles gives an adequate numerical precision  for most applications.

\section{Simulation study\label{subsec:Simulations}}

We perform several simulation experiments to study the
performance of the proposed inference methodology on data
generated from both static and dynamic data generating processes (DGPs).

\subsection{Simulation experiments \label{subsec:simulation experiments}}

The simulation experiments simulate data from the five data generating processes summarized in Table \ref{table:DGPsimulation}. Models $\mathrm{M}_{1}$, $\mathrm{M}_{3}$, and $\mathrm{M}_{3}$ all assume a batch data structure, which is common in industrial applications, where data are observed at irregular time points and aggregated into batches.  The only source of time variation is in the parameter evolution; the parameter stays constant within a batch but may change across batches. Models $\mathrm{M}_{4}$ and $\mathrm{M}_{5}$ are mixtures of autoregressive experts \citep{carvalho2005modeling, carvalho2007modelling} with constant ($\mathrm{M}_{4}$) and time-varying parameters ($\mathrm{M}_{5}$). The latter two models are pure time series models with one data point observed at equidistant times, and where the response depends on its lagged values. We follow \citet{carvalho2007modelling} and use $\log(y_{t-1} +1)$ as lagged values, and parametrizing the autoregressive parameter as $1-\exp(\beta)$, so that $\beta$ can vary freely while at the same time ensuring that the process is stationary for every parameter value.

For each DGP, $50$ datasets of $1000$ observations are generated. For $\mathrm{M}_{1}$ to $\mathrm{M}_{3}$, data are generated sequentially over $10$ time intervals, having $100$ observations within each interval. For $\mathrm{M}_{4}$ and $\mathrm{M}_{5}$ a time series of length $1000$ is generated. The first half of the data is used for training and the last for validation;  all model comparisons are based on LPS values computed on the test set - the last half of generated data.

\begin{table}[H]
\noindent \centering{}\caption{{\small{}All data generating processes used in the simulation
study. For $\mathrm{M}_{1}$, $\mathrm{M}_{2}$ and $\mathrm{M}_{3}$ the covariates $x_{ij}$ and $z_{ij}$ are iid $U(-1,1)$.\medskip{}
}}
\begin{tabular}{l}
\hline 
\tabularnewline
Model $\mathrm{M}_{1}$ - \textbf{Static Poisson regression}\tabularnewline
 \qquad  \qquad $y_{ij}|\mathbf{x}_{ij}\sim\mathrm{Po}\left(\lambda_{ij}\right),$\tabularnewline 
 \qquad  \qquad $\log\lambda_{ij}=\mathbf{x}_{ij}^{\prime}\boldsymbol{\beta}$,
 \qquad $\boldsymbol{\beta}=(1,\log(0.5))$\tabularnewline
\tabularnewline
Model $\mathrm{M}_{2}$ - \textbf{Dynamic Poisson regression}\tabularnewline
 \qquad  \qquad $y_{ij}|\mathbf{x}_{ij}\sim Po\left(\lambda_{ij}\right)$\tabularnewline
 \qquad  \qquad $\log\lambda_{ij}\mathbf{=x}_{ij}^{\prime}\boldsymbol{\beta}_{j}$,\tabularnewline 
\tabularnewline
$\boldsymbol{\beta}_{j}=\boldsymbol{\beta}_{j-1}+u_{j}$, 
\qquad $u_{j}\sim\mathrm{N}\left(0,U\right)$, \tabularnewline 
$\boldsymbol{\beta}_{0}=(1,\log(0.5))$,
 \qquad $U=\textrm{Diag}(.04^2,\,.04^2)$\tabularnewline
\tabularnewline
Model $\mathrm{M}_{3}$ - \textbf{Dynamic mixture of Poisson regression experts}\tabularnewline
 \qquad \qquad $y_{ij}|\mathbf{x}_{ij},\mathbf{z}_{ij}\sim\sum_{k=1}^{2}\phi_{ijk}\mathrm{Po}\left(\lambda_{ijk}\right)$\tabularnewline
 \qquad  \qquad $\log\lambda_{ijk}=\mathbf{x}_{ij}^{\prime}\boldsymbol{\beta}_{jk}$,\qquad 
 
 $\phi_{ij,2}=\textrm{logit}\left(\mathbf{z}_{ij}^{\prime}\boldsymbol{\theta}_{j}\right)$, \tabularnewline 
 \tabularnewline
 $\boldsymbol{\beta}_{jk}=\boldsymbol{\beta}_{j-1,k}+u_{jk}$,
 \qquad $u_{jk}\sim\mathrm{N}\left(0,\,U_{k}\right)$, \tabularnewline 
  $\boldsymbol{\beta}_{1,0}=(1,\log(0.5))$,
 \qquad $\boldsymbol{\beta}_{2,0}=(-2,\log(0.5))$,\tabularnewline 
 \tabularnewline 
 $\boldsymbol{\theta}_{j}=\boldsymbol{\theta}_{j-1}+v_{j}$,
  \qquad$v_{j}\sim\mathrm{N}\left(0,\,V\right)$,
  \tabularnewline 
 $\boldsymbol{\theta}_{0}=(2,-1)$,
 \qquad $U_{1}=U_{2}=V=\textrm{Diag}(0.04^2,0.04^2)$\tabularnewline
\tabularnewline
Model $\mathrm{M}_{4}$ - \textbf{Static mixture of Poisson autoregressive experts}\tabularnewline
\qquad \qquad $y_{j}|y_{j-1}\sim\sum_{k=1}^{2}\phi_{jk}\mathrm{Po}\left(\lambda_{jk}\right)$\tabularnewline
\qquad \qquad $\log\lambda_{jk}=\beta_{0k} + \beta_{1k}\log(y_{j-1}+1)$,\tabularnewline
\qquad \qquad $\phi_{j,2}=\textrm{logit}\left(\theta_{0} + \theta_{1}\log(y_{j-1}+1)\right)$\tabularnewline
\tabularnewline
$\boldsymbol{\beta}_{1}=(1,0.5$),
\qquad $\boldsymbol{\beta}_{2}=(-2,0.5)$,
\qquad $\boldsymbol{\theta}=(2,-1)$\tabularnewline
\tabularnewline
Model $\mathrm{M}_{5}$ - \textbf{Dynamic mixture of Poisson autoregressive experts}\tabularnewline
\qquad \qquad$y_{j}|y_{j-1}\sim\sum_{k=1}^{2}\phi_{jk}\mathrm{Po}\left(\lambda_{jk}\right)$\tabularnewline
\qquad \qquad $\log\lambda_{jk}=\beta_{0jk} + (1 - \exp(\beta_{1jk}))\log(y_{j-1}+1)$, \tabularnewline
\qquad \qquad $\phi_{j,2}=\textrm{logit}\left(\theta_{0j} + \theta_{1j}\log(y_{j-1}+1)\right)$, \tabularnewline
\tabularnewline
$\boldsymbol{\beta}_{jk}=\boldsymbol{\beta}_{j-1,k}+u_{jk}$, 
\qquad $u_{jk}\sim\mathrm{N}\left(0,\,U_{k}\right)$,
\tabularnewline 
$\boldsymbol{\beta}_{0,1}=(1,\log(0.5)$), 
\qquad$\boldsymbol{\beta}_{0,2}=(-2,\log(0.5))$,\tabularnewline
\tabularnewline 
$\boldsymbol{\theta}_{j}=\boldsymbol{\theta}_{j-1}+v_{j}$,
\qquad $v_{j}\sim\mathrm{N}\left(0,\,V\right),$
\tabularnewline 
$\boldsymbol{\theta}_{0}=(2,-1)$,
\qquad $U_{1}=U_{2}=V=\textrm{Diag}(0.04^2,0.025^2)$ \tabularnewline
\tabularnewline
\hline 
\end{tabular}\label{table:DGPsimulation}
\end{table}

\subsection{Inference of the number of mixture components and the discount factor \label{subsec:discount factor_inference}}

\begin{figure}[H]
\begin{centering}
\includegraphics[scale=0.6]{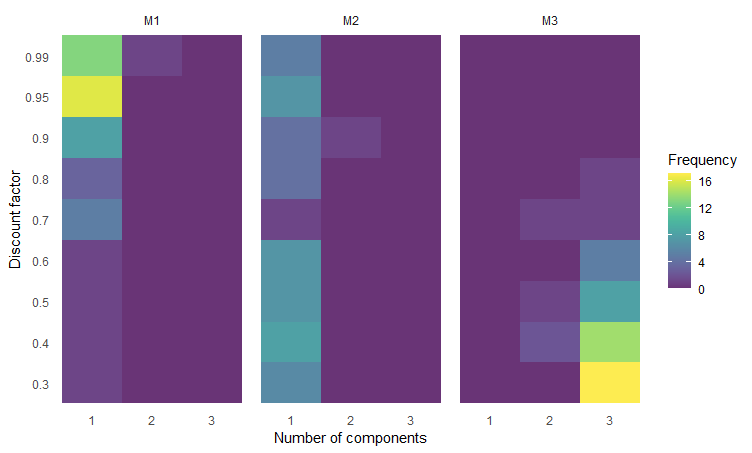}
\par\end{centering}
\caption{{\small{}Fitting different models to the data generated from the three $\mathrm{M}_{1}$ - $\mathrm{M}_{3}$
DGPs. Each panel displays the number of times each model was selected  based on the LPS. Results are based on $1000$ particles.}{\footnotesize{} \label{fig:Model selection frequency}}}
\end{figure}

The number of mixture components/experts $K$ and the discount factor $\alpha$ discussed in Section \ref{sec:The model} are unknown. Inference of these quantities is a research area. Here, we use LPS to assess the performance of the proposed methodology on the inference of these quantities.  Several models  with $k=1, 2, 3$ Poisson components and $\alpha\in\{0.3, 0.4, 0.5, 0.6, 0.7, 0.8, 0.9, 0.99\}$ are fitted to data generated from each of the $\mathrm{M}_{1}$, $\mathrm{M}_{2}$ and $\mathrm{M}_{3}$ DGPs, and  LPS is used to select the best model. The aim is to see if the proposed inference methodology is able to identify the underlying data generating process. Figure \ref{fig:Model selection frequency} displays the selection frequency of $K$ and $\alpha$ for all fitted models. 

For $\mathrm{M}_{1}$ and $\mathrm{M}_{2}$, the most frequently selected model is the single component Poisson model with $\alpha=0.99$, and $\alpha=0.4$ respectively. While, for $\mathrm{M}_{3}$, it is the model with $\alpha=0.6$, and not the correct two components mixture model. This slight overestimation is not surprising as LPS is often observed to have a tendency to be generous with the number of components in a mixture without having a large impact on the final predictive density, see e.g. \citet{villani2012generalized}. 
\subsection{Comparing static and dynamic models \label{subsec:static_dynamic_models_comparison}}

The data generating process is generally unknown in real applications and the usual strategy in modeling the data is to fit static models. It is therefore interesting to evaluate how fitting a dynamic model would differ from its static version in the cases where  the true data generating process is static or dynamic. 

\begin{figure}[H]
\begin{centering}
\includegraphics[scale=.6]{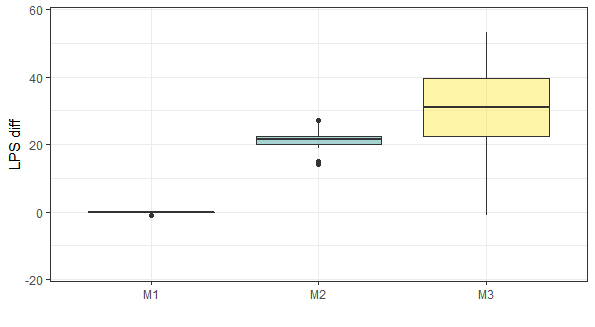}
\par\end{centering}
\caption{{\small{}Boxplot of the difference in LPS of the model selected in the validation step and the corresponding static model for the first three data generating processes. }\label{fig:Comparison-of-the selected and static models_M1M3}}
\end{figure}

We first consider the $\mathrm{M_1}$, $\mathrm{M_2}$ and $\mathrm{M_3}$ data generating processes. Figure \ref{fig:Comparison-of-the selected and static models_M1M3} compares the performance of the model i) with $K = K_{opt}$ and $\alpha = \alpha_{opt}$, where $K_{opt}$
and $\alpha_{opt}$ are the values chosen from LPS and ii) the corresponding static model with $K = K_{opt}$ and
$\alpha = 0.99$. The figure shows boxplots of the difference in the LPS values in the validation set for
both models. For $\mathrm{M_1}$ the average LPS difference between the selected and the static models is
around zero, which shows that the dynamic model does not overfit on static data. On the other
hand, for the two dynamic data generating processes, $\mathrm{M_2}$ and $\mathrm{M_3}$, the dynamic model selected in
the validation step clearly outperforms the static model and the difference in LPS increases with
the number of components.

Consider now the mixture of autoregressive Poisson experts model, $\mathrm{M}_4$ and $\mathrm{M}_5$. Static models are fitted using the MCMC algorithm in \citet{villani2012generalized} with $10000$ MCMC iterations and dynamic models are fitted using $1000$ particles. We partition the data into batches when running our algorithm, which also allows us to investigate the effect of the chosen batch size. Three different batch sizes are compared: $10$, $25$ and $50$. The MCMC algorithm is also updated sequentially at each batch for comparability and for reducing computing times. The dynamic models are trained with different discount factors $\alpha\in\{.3,.4,.5,.6,.7,.8,.9,.99\}$ and the LPS is used to select the best model in the validation step. 

\begin{figure}[H]
\begin{centering}
\includegraphics[scale=.6]{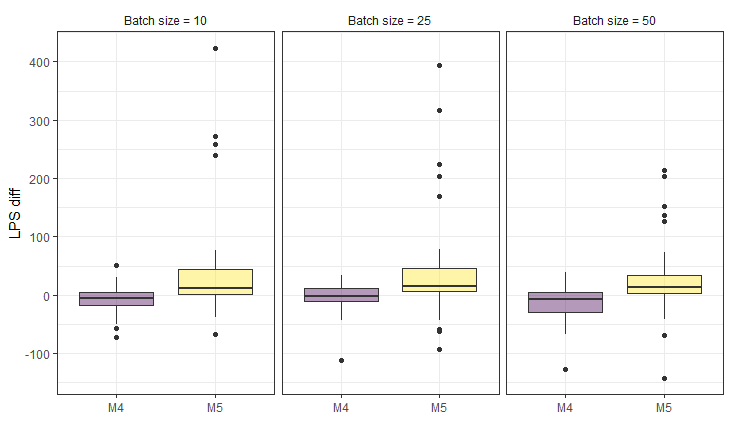}
\par\end{centering}
\caption{{\small{}Boxplot of the difference in LPS of the model selected the dynamic model and the corresponding static model. }\label{fig:Comparison-of-the selected and static models}}
\end{figure}

Figure \ref{fig:Comparison-of-the selected and static models} displays boxplots of the LPS difference of the selected dynamic model and the static model.
The average LPS difference between the selected dynamic model and the static model is around zero, which again shows that the dynamic model does not overfit on static data. On the other hand, for the two dynamic data generating processes, the dynamic model clearly outperforms the static model. Also, one can note that the data partition has minimal effect as there is not much variation in the LPS differences for the different batch partitions.

\subsection{Efficiency of the linear Bayes proposal \label{subsec:proposal_efficiency}}

The efficiency of particle filter algorithms is generally assessed based on the effective sample size
\[
\textrm{ESS}_{j}:=\frac{1}{\sum_{m=1}^{M}(w_{j}^{m})^{2}},
\]
where $w_{j}^{m}$ are the importance weights computed at interval $j$. 

 To assess the performance of the linear Bayes proposal strategy, we compare it with the local linearisation proposal strategy \citep{doucet2000sequential}; the state-of-the-art method of constructing proposal densities which approximates the target density (\ref{eq:Filtering distribution-1}) by  a linear Gaussian distribution obtained from a second order Taylor expansion of the target density with respect to the regression coefficients $\gamma_{j}$, for $j \in (1,\cdots,J)$.  The main difference between these two methods is that the local linearisation method approximates the target density by a Gaussian density without the intermediate step of updating the linear predictors. 

\begin{figure}[H]
\begin{centering}
\includegraphics[scale=0.65]{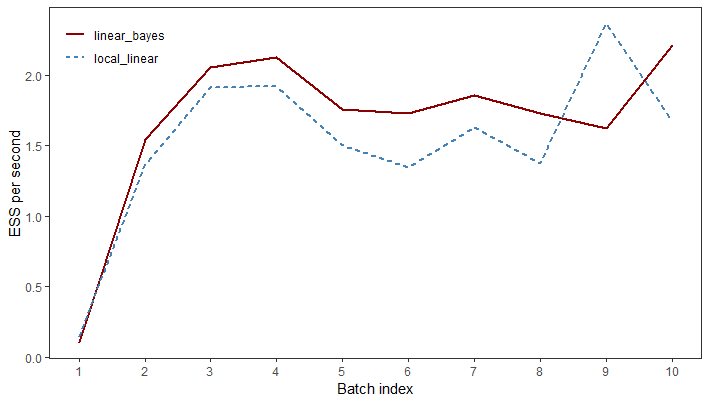}
\par\end{centering}
\caption{{\small{}Comparing the effective sample size per second for the linear Bayes and local linear proposal strategies.}\label{fig:Comparison ess rate}} 
\end{figure}

Figure \ref{fig:Comparison ess rate} compares the effective sample size per second generated by the two strategies as a way of comparing their efficiency and computation time. The results are based on data simulated from $\mathrm{M}_3$ and a posterior distribution approximated by $1000$ particles and $\alpha=0.5$. 
The figure shows that the linear Bayes proposal generates an effective sample size that is on average $10\%$ higher than the local linearisation. Both methods are quite fast; their computation time on a simple windows laptop with intel core $i5$ processor is less than $3$ CPU minutes.

\section{Conclusions}

We introduce a general class of dynamic mixture of experts models
for online predictions; the model allows the regression coefficients
in each mixture component and weight to vary over time. The component
models can be essentially any density function, not necessarily limited
to the exponential family.

We propose an efficient SMC algorithm for sequential inference and
online prediction that is tailored to handle the proposed model class
with potentially high-dimensional parameter spaces. The algorithm
handles models with static and dynamic parameters in a unified way.

The model is applied to online prediction of the number of faults
in a continuously upgraded large-scale industrial software project.
We show that allowing the parameters to evolve over time greatly improves
the model's predictive performance. A simulation study documents that
the proposed model selection procedure is i) effective in reducing
flexibility when data comes from a static single-component model,
ii) able to fit data from multi-component models with time-varying
parameters, and iii) it is fast and generates an effective sample size rate that is superior to the state-of-the-art particle filter which uses a proposal density designed via the local linearization of the target density.

\bibliographystyle{chicago}
\phantomsection\addcontentsline{toc}{section}{\refname}\bibliography{bibliography}

\appendix
\section{Gradient and Hessian for dynamic
mixture of experts models}\label{sec:gradient_hessian_details}

\subsection{Gradient and Hessian for general dynamic
mixture of experts models \label{sec:gradient_hessian_general} }

This section provides details on the expressions of the gradient and the  Hessian  of the model \eqref{eq:Dynamic model} discussed in Section \ref{subsec:Proposal-distribution-based-1}.  Let $\pi_{jk}=\log\omega_{jk}f_{jk}\left(y_{j}|\lambda_{jk}\right)$
as in Section \ref{subsec:Proposal-distribution-based-1}, the gradient 
\begin{align*}
\nabla_{\boldsymbol{\rho}_{j}}\log p(\boldsymbol{\rho}_{j}\vert D_{1:j}) & =\nabla_{\boldsymbol{\rho}_{j}}\log\left(\sum_{k=1}^{K}\exp\left(\pi_{jk}\right)\right)-\Sigma_{\rho_{j}}^{-1}(\boldsymbol{\rho}_{j}-\bar{\boldsymbol{\rho}}_{j})\\
 & =\sum_{k=1}^{K}\frac{\nabla_{\boldsymbol{\rho}_{j}}\exp\left(\pi_{jk}\right)}{\sum_{k=1}^{K}\exp\left(\pi_{jk}\right)}-\Sigma_{\rho_{j}}^{-1}(\boldsymbol{\rho}_{j}-\bar{\boldsymbol{\rho}}_{j}).
\end{align*}

since $\nabla_{\boldsymbol{\rho}_{j}}\exp\left(\pi_{jk}\right)=\exp\left(\pi_{jk}\right)\nabla_{\boldsymbol{\rho}_{j}}\pi_{jk}$,
we have
\[
\nabla_{\boldsymbol{\rho}_{j}}\log p(\boldsymbol{\rho}_{j}\vert D_{1:j})=\sum_{k=1}^{K}\frac{\exp\left(\pi_{jk}\right)}{\sum_{k=1}^{K}\exp\left(\pi_{jk}\right)}\nabla_{\boldsymbol{\rho}_{j}}\pi_{jk}-\Sigma_{\rho_{j}}^{-1}(\boldsymbol{\rho}_{j}-\bar{\boldsymbol{\rho}}_{j}),
\]
Hence,
\[
\nabla_{\boldsymbol{\rho}_{j}}\log p(\boldsymbol{\rho}_{j}\vert D_{1:j})=\sum_{k=1}^{K}\mathrm{Pr}(s_{j}=k\vert D_{1:j})\nabla_{\boldsymbol{\rho}_{j}}\pi_{jk}-\Sigma_{\rho_{j}}^{-1}(\boldsymbol{\rho}_{j}-\bar{\boldsymbol{\rho}}_{j}).
\]
The Hessian is 
\begin{align*}
\nabla\nabla_{\boldsymbol{\rho}_{j}}\log p(\boldsymbol{\rho}_{j}\vert D_{1:j}) & =\sum_{k=1}^{K}\left[\nabla_{\boldsymbol{\rho}_{j}}\mathrm{Pr}(s_{j}=k\vert D_{1:j})\right]\nabla_{\boldsymbol{\rho}_{j}}\pi_{jk}+\sum_{k=1}^{K}\mathrm{Pr}(s_{j}=k\vert D_{1:j})\nabla\nabla_{\boldsymbol{\rho}_{j}}\pi_{jk}-\Sigma_{\rho_{j}}^{-1},
\end{align*}
where
\begin{align*}
\nabla_{\boldsymbol{\rho}_{j}}\mathrm{Pr}(s_{j} & =k\vert D_{1:j})=\nabla_{\boldsymbol{\rho}_{j}}\frac{\exp\left(\pi_{jk}\right)}{\sum_{h=1}^{K}\exp\left(\pi_{jh}\right)}\\
 & =\frac{\left[\sum_{h=1}^{K}\exp\left(\pi_{jh}\right)\right]\nabla_{\boldsymbol{\rho}_{j}}\exp\left(\pi_{jk}\right)-\exp\left(\pi_{jk}\right)\left[\sum_{h=1}^{K}\nabla_{\boldsymbol{\rho}_{j}}\exp\left(\pi_{jh}\right)\right]}{\left[\sum_{h=1}^{K}\exp\left(\pi_{jh}\right)\right]^{2}}\\
 & =\frac{\exp\left(\pi_{jk}\right)}{\sum_{h=1}^{K}\exp\left(\pi_{jh}\right)}\left[\nabla_{\boldsymbol{\rho}_{j}}\pi_{jk}-\sum_{h=1}^{K}\frac{\exp\left(\pi_{jh}\right)}{\sum_{h=1}^{K}\exp\left(\pi_{jh}\right)}\nabla_{\boldsymbol{\rho}_{j}}\pi_{jh}\right]\\
 & =\mathrm{Pr}(s_{j}=k\vert D_{1:j})\nabla_{\boldsymbol{\rho}_{j}}\pi_{jk}-\sum_{h=1}^{K}\mathrm{Pr}(s_{j}=h\vert D_{1:j})\nabla_{\boldsymbol{\rho}_{j}}\pi_{jh}.\\
 & =\sum_{k=1}^{K}\mathrm{Pr}(s_{j}=k\vert D_{1:j})\left[\nabla_{\boldsymbol{\rho}_{j}}\pi_{jk}-\sum_{h=1}^{K}\mathrm{Pr}(s_{j}=h\vert D_{1:j})\nabla_{\boldsymbol{\rho}_{j}}\pi_{jh}\right]\nabla_{\boldsymbol{\rho}_{j}}\pi_{jk}\\
 & +\sum_{k=1}^{K}\mathrm{Pr}(s_{j}=k\vert D_{1:j})\nabla\nabla_{\boldsymbol{\rho}_{j}}\pi_{jk}-\Sigma_{\rho_{j}}^{-1}.
\end{align*}
It can easily be shown that the first term in the expression above
is zero; hence 
\begin{align*}
\nabla\nabla_{\boldsymbol{\rho}_{j}}\log p(\boldsymbol{\rho}_{j}\vert D_{1:j}) & =\sum_{k=1}^{K}\mathrm{Pr}(s_{j}=k\vert D_{1:j})\nabla\nabla_{\boldsymbol{\rho}_{j}}\pi_{jk}-\Sigma_{\rho_{j}}^{-1}.
\end{align*}

\subsection{Poisson experts}\label{sec:Examples-of-component}

This appendix provides details on the mixture of experts model with
Poisson components fitted to the software reports data. The mixture of experts model with Poisson components for the batch
$D_{j}$, $j=1,\ldots,21$, has the form 
\[
f_{j}(y_{ij}|\mathbf{\tilde{x}}_{ij},\boldsymbol{\omega}_{ij},\boldsymbol{\lambda}_{ij})=\sum_{k=1}^{K}\omega_{ijk}\frac{1}{y_{ik}!}\lambda_{ijk}^{y_{ij}}\exp\left\{ -\lambda_{ijk}\right\} ,k=1,\ldots,K
\]

\[
\omega_{ijk}=\frac{\exp\left(\psi_{ijk}\right)}{1+\sum_{j=2}^{J}\exp\left(\psi_{ijk}\right)}
\]

\[
\eta_{ijk}=\log\left(\lambda_{ijk}\right)=\mathbf{x}_{ij}^{\prime}\boldsymbol{\beta}_{jk},\,\,\psi_{ijk}=\mathbf{z}_{ij}^{'}\boldsymbol{\theta}_{jk}.
\]

To compute the gradient and Hessian required in the proposal density, we define (omitting the index $i$)
\begin{align*}
l_{k} & =y_{j}\log\lambda_{jk}-\lambda_{jk}+\psi_{jk}-\log\left(1+\sum_{k=2}^{K}\exp\left(\psi_{jk}\right)\right),
\end{align*}

The first derivatives are
\[
\frac{\partial l_{k}}{\partial\eta_{jk}}=y_{j}-\exp\left(\eta_{jk}\right),\,\,\frac{\partial l_{k}}{\partial\psi_{jk}}=1-\frac{\exp\left(\psi_{jk}\right)}{1+\sum_{k=2}^{K}\exp\left(\psi_{jk}\right)}
\]

and the second derivatives are

\[
\frac{\partial^{2}l_{k}}{\partial\eta_{jk^{2}}}=-\exp\left(\eta_{jk}\right),\,\,\,\frac{\partial^{2}l_{k}}{\partial\psi_{jk}^{2}}=-\frac{\exp\left(\psi_{jk}\right)}{1+\sum_{k=2}^{K}\exp\left(\psi_{jk}\right)}\left[1-\frac{\exp\left(\psi_{jk}\right)}{1+\sum_{k=2}^{K}\exp\left(\psi_{jk}\right)}\right]
\]

\[
\frac{\partial^{2}l_{k}}{\partial\psi_{jk}\partial\psi_{jh}}=\frac{\exp\left(\psi_{jk}+\psi_{jh}\right)}{\left(1+\sum_{k=2}^{K}\exp\left(\psi_{jk}\right)\right)^{2}},\,\,\frac{\partial^{2}l_{k}}{\partial\eta_{jk}\partial\eta_{jh}}=\frac{\partial^{2}l_{k}}{\partial\psi_{jk}\partial\eta_{jh}}=0\,\,h\neq k
\]

\subsection{Generalized Poisson experts \label{sec:generalized poisson details}}
This section delineates the one-component dynamic generalized Poisson model fitted in Section \ref{subsec:Software-trouble-reports}. To simplify the notation we omit the batch index $j$. The generalized Poisson model is of the form \citep{famoye2006zero} 
\begin{equation}
    f(y \vert \mu, \varphi) = \Bigg(\frac{\mu}{1+\varphi\mu} \Bigg)^y \frac{(1+\varphi y)^{y-1}}{y!} \exp\Bigg(\frac{-\mu(1+\varphi y)}{1+\varphi\mu} \Bigg),
\end{equation}
where $\mu>0$ is the mean and $\varphi>0$ is the overdispersion parameter. The mean and the dispersion parameters  are connected to covariates via a log links $\mu=\exp(\boldsymbol{x}^\top\boldsymbol{\beta})$ and $\varphi=\exp(\boldsymbol{x}^\top\boldsymbol{\theta})$ respectively.

Let $l = \log f(y\vert \mu, \varphi) $, the first derivatives are

\[
\frac{\partial l}{\partial\mu}=\frac{y-\mu}{\mu\left(1+\varphi\mu\right)^{2}},
\]
\[
\frac{\partial l}{\partial\varphi}=\frac{\left(y^{2}-y\right)\left(1+\varphi\mu\right)-\mu\left(1+\varphi y\right)\left[y\left(1+\varphi\mu\right)+\left(y-\mu\right)\right]}{\left(1+\varphi y\right)\left(1+\varphi\mu\right)^{2}},
\]

and the second derivatives are 
\[
\frac{\partial^{2}l}{\partial\mu^{2}}=\frac{y\left(1+\varphi\mu\right)+2\left(y-\mu\right)\varphi\mu}{\mu^{2}\left(1+\varphi\mu\right)^{3}},
\]
\[
\frac{\partial^{2}l}{\partial\mu\partial\varphi}=\frac{2\left(y-\mu\right)}{\mu\left(1+\varphi\mu\right)^{3}}
\]

\[
\frac{\partial^{2}l}{\partial\varphi^{2}}=\frac{\left(y^{3}-y^{2}\right)\left(1+\varphi\mu\right)^{3}-\mu^{2}\left(1+\varphi y\right)^{2}\left[y\left(1+\varphi\mu\right)+\left(y-\mu\right)\right]}{\left(1+\varphi y\right)^{2}\left(1+\varphi\mu\right)^{3}}
\]

\section{Identifiability}\label{sec:identifiability}
\citet{jiang1999identifiability} prove that mixtures of generalized linear model experts are identifiable if four conditions hold: i) the experts are irreducible, i.e. no pair of experts have identical parameters; ii) there is an ordering of the parameters to avoid 
so-called label switching of components; iii) the parameters in mixing functions are set to zero for one of the components; and iv) a certain nondegeneracy condition holds that precludes exact linear combinations of the expert densities. The first condition can be assumed to always hold, otherwise we can just collapse identical experts to a single one and reduce the number of mixture components \citep{jiang1999identifiability}. Condition ii) is rarely explicitly imposed in mixture models since it complicates inference, and unrestricted inference will therefore returns one of the $K!$ identical modes; this is acknowledged in \citet{jiang1999identifiability} who recommend in Remark 1 to report the mode corresponding to the order restricted parameters for interpretation. Condition iii) is explicitly imposed in our models by zero restrictions, as is commonly done in mixture of experts models. Condition iv) is a technical condition that should be checked for each distributional family on a case by case basis, but is instead often silently assumed to hold. \citet{jiang1999identifiability} prove that several commonly used distributions satisfy the nondegeneracy condition, including the Poisson. In fact, only for binomial experts with the number trials smaller than $2K-1$ do \citet{jiang1999identifiability} find that the nondegeneracy does not hold, and even then they conjecture that it will hold "for almost all parameters".

\citet{jiang1999identifiability} prove their results for generalized linear regression (GLM) components, i.e. for densities in the one-parameter exponential family with a scalar dispersion parameter. However, their proofs only rely on properties of the translation and permutation groups acting on densities, and do not use specific properties of exponential families. Their results therefore also apply to a mixture of GLM-type  experts with a single parameter depending covariates, but with a density that may be outside of the exponential family; an example of such a model is the generalized Poisson regression where the mean depends on covariates via a linear predictor through a link function, and the overdispersion parameter is a constant. We will now show that this model satisfies the nondegeneracy condition and a mixture of such experts is therefore identified. For the more general model used in \citet{villani2012generalized} and in Section \ref{subsec:Software-trouble-reports}, where  the overdispersion parameter is also allowed to depend on covariates, the issue of identification is not yet resolved. We conjecture however that this model is also identified based on the discussion in the previous paragraph and on the empirical results in \citet{villani2012generalized} where no convergence issues were encountered when using that model.

To prove that a mixture of generalized Poisson regression experts satisfies the nondegeneracy condition we follow the same technique as in the proof of Lemma 3(a) in \citet{jiang1999identifiability} for the identifiability of mixture of Poisson experts. The nondegeneracy condition (Condition 1 in \citet{jiang1999identifiability}) is that $\{f(y \vert \mu_k,\varphi_k) \}_{k=1}^{2K}$ are $2K$ linearly independent functions of $y$ for any $2K$ distinct pairs $(\mu_k,\varphi_k),\text{ for } k=1,\ldots,2K$, i.e. that 
\begin{equation}\label{jiangtannercondition}
    \sum_{k=1}^{2K} c_k f(y \vert \mu_k,\varphi_k) = 0 \text{ for all } y \in \{0,1,\ldots\},
\end{equation}
only for $c_1=c_2=\ldots=c_{2K}=0$.
Now, if $\sum_{k=1}^{2K} c_k f(y \vert \mu_k,\varphi_k) = 0 $ then
\begin{equation}
    y!y^{1-y}\sum_{k=1}^{2K} c_k  f(y \vert \mu_k,\varphi_k) = \sum_{k=1}^{2K} c_k (y^{-1}+\varphi_k)^{y-1} \Bigg(\frac{\mu_k}{1+\varphi_k\mu_k} \Bigg)^y \exp\Bigg(\frac{-\mu_k(1+\varphi_ky)}{1+\varphi_k\mu_k} \Bigg) =  0. 
\end{equation}
We can write 
\begin{equation}\label{eq:jiangtannerres}
    (y^{-1}+\varphi_k)^{y-1}\Bigg(\frac{\mu_k}{1+\varphi_k\mu_k} \Bigg)^y \exp\Bigg(\frac{-\mu_k(1+\varphi_ky)}{1+\varphi_k\mu_k} \Bigg) = \exp\big(a_k + b_k y + (y-1)\log(y^{-1}+\varphi_k)\big),
\end{equation}
where $a_k = -\mu_k/(1+\varphi_k\mu_k)$ and $b_k = \log(\mu_k/(1+\varphi_k\mu_k)) - \mu_k\varphi_k/(1+\varphi_k\mu_k)$. Now, as $y\rightarrow \infty$ the expression in \eqref{eq:jiangtannerres} behaves as $\exp(\tilde{a}_k + \tilde{b}_k y)$  where $\tilde{a}_k=a_k-\log\varphi_k$ and $\tilde{b}_k = b_k+\log\varphi_k$. Similar to the proof of Lemma 3(a) in \citet{jiang1999identifiability} we therefore have that as $y\rightarrow \infty$ the decay/explosion rates of $\exp(\tilde{a}_k + \tilde{b}_k y)$ are different for each $k$ since the $\mu_k$ and/or $\varphi_k$ are different. Hence since the equality in \eqref{jiangtannercondition} needs to hold for all $y \in \{0,1,\ldots\}$, this can only be true for $c_1=c_2=\ldots=c_{2K}=0$. The mixture of generalized poisson experts therefore satifies the nondegeneracy condition.
\end{document}